\documentstyle[aps,twocolumn,prl,epsf]{revtex}

\newcommand{\be}{\begin{equation}}
\newcommand{\ee}{\end{equation}}
\newcommand{\bea}{\begin{eqnarray}}
\newcommand{\eea}{\end{eqnarray}}

\def\b6{CeB$_6$}

\begin{document}
\draft

 \twocolumn[\hsize\textwidth\columnwidth\hsize\csname @twocolumnfalse\endcsname
\title{Two-Channel Kondo Lattice: An Incoherent Metal}
\author{
Mark Jarrell$^{(a)}$, Hanbin Pang$^{(a)}$, D.L.\ Cox$^{(b,c)}$, and
K.H.\ Luk$^{(d)}$.
}
\address{
$^{(a)}$ Department of Physics,
University of Cincinnati, Cincinnati, OH 45221\\
$^{(b)}$Department of Physics,
The Ohio State University, Columbus, OH, 43202 \\
$^{(c)}$ Institute for Theoretical Physics, University of California,
Santa Barbara, CA 93106-4030\\
$^{(d)}$ Research Centre, The Hong
Kong University of Science and Technology, Clear Water Bay, Kowloon,
Hong Kong \\
}
\date{\today}
\maketitle

\widetext
\begin{abstract}
\noindent
The two-channel Kondo lattice model is examined with a Quantum Monte
Carlo
simulation in the limit of infinite dimensions.   We find
non-fermi-liquid
behavior at low temperatures including a finite low-temperature
single-particle
scattering rate, the lack of a fermi edge and Drude weight.  However,
the low-energy density of electronic states is finite.  Thus, we
identify this
system as an incoherent metal.  We discuss the relevance of our results
for concentrated heavy fermion metals with non-Fermi-Liquid behavior.
\end{abstract}

 ] 

\narrowtext
\paragraph*{Introduction.}
The Fermi Liquid theory of Landau has provided a remarkably robust
paradigm
for describing the properties of interacting fermion systems such as
liquid $^3$He and numerous metals like aluminum and copper. The key
notion of this theory is that the low lying excitations of the
interacting system possess a 1:1 map to those of the noninteracting
system and hence are called ``quasiparticles.''  In the metallic
context, one may think of the quasiparticles as propagating
electron-like
wave packets with renormalized magnetic moment and effective mass
reflecting the ``molecular field'' of the surrounding medium.
A sharp Fermi surface remains in the electron occupancy function
$n_{\vec k}$ which measures the number of electrons with a given
momentum,
and for energies $ \omega$ and temperatures $T$
asymptotically close to the Fermi surface the excitations have a
decay rate going as $\omega^2 + \pi^2 (k_BT)^2$, which is much smaller
than the quasiparticle energy, which generally translates into a $T^2$
contribution to the electrical resistivity $\rho(T)$.
This theory has proven
useful in describing phase transitions within the Fermi liquid, such as
superconductivity which is viewed as a pairing of Landau quasiparticles
in conventional metals such as Al.
 
  
The Fermi liquid paradigm appears now to be breaking down empirically
in numerous  materials, notably the quasi-two-dimensional cuprate
superconductors\cite{cupnfl} and a number of fully three dimensional
heavy fermion alloys and
compounds\cite{hfnfl}.  In these systems such anomalies as a
conductivity
with linear dependence on $\omega,T$ are often observed, as are
logarithmically
divergent linear specific heat coefficients.  If the quasiparticle
paradigm indeed breaks down, this may require completely new concepts
to explain the superconducting phases of these materials.
While the Luttinger liquid
theory provides an elegant way to achieve non-Fermi liquid theory in
one-dimension (with, e.g., no jump discontinuity in $n_{\vec k}$, and 
separation or unbinding of spin and charge quantum numbers),
this results from the special point character of the
Fermi surface. Whether the essential spin-charge
separation may ``bootstrap'' 
into higher dimensions remains unclear\cite{lutnfl}.
Among
the remaining theories to explain experiment are those based upon
proximity to a zero temperature quantum critical point\cite{qcpnfl},
those based upon disorder induced distributions of Kondo scales in local
moment systems\cite{disknfl}, and those which hope to explain the
physics from impurity to lattice crossover effects in the multi-channel
Kondo model\cite{twochnfl}.  Notably lacking for dimensions higher than
one are rigorous solutions to microscopic models which display non-Fermi
liquid behavior\cite{butkotsi}.
 
In this Letter, we present the first rigorous solution of the
two-channel Kondo lattice model in infinite spatial dimensions.  We
find that the paramagnetic phase of this model is an ``incoherent metal''
with finite density of states at the Fermi energy and finite residual
resistivity.  The excitation spectrum is non-fermi liquid like; in
particular, there is no discontinuity in $n_{\vec k}$, a finite lifetime
for electrons at the Fermi energy, an ill defined quasiparticle mass,
and a linear in temperature saturation of the electrical resistivity.
We find that physical quantities may be suitably scaled with a lattice
Kondo scale $T_0$ that is significantly enhanced over the impurity
limit.  We discuss the possible relevance of these results to understand
transport properties of concentrated heavy electron materials.
  

\paragraph*{Motivation} The two-channel Kondo impurity model consists of
two identical species of non-interacting electrons antiferromagnetically
coupled to a spin 1/2 impurity.  Non-fermi liquid behavior results
because of the inability to remove the impurity spin:  it is
energetically favorable for both conduction electron bands to couple to the 
impurity which gives a spin 1/2 complex on all length scales.  As a
result, the ground state is degenerate and the excitation spectrum
non-Fermi liquid like.  In contrast, the single channel Kondo model has
a singlet ground state with the impurity spin screened out, and a Fermi
liquid excitation spectrum corresponding to the removal of one 
conduction state from the system.    On extension to the lattice and
ignoring the renormalization of the environment around each spin, 
the array of single channel model 
singlets would simply renormalize the potential scattering. 
In contrast, the array of many body doublets in the two-channel case would give
rise to a dynamical spin-disorder scattering in the absence of any
cooperative transition that lent coherence to the spin array.  We might
thus anticipate a finite residual resistivity and other non-Fermi liquid
properties in the paramagnetic phase\cite{coxinfd}.

\paragraph*{Model} The Hamiltonian for the two-channel Kondo
lattice is 
$$H=J\sum_{i,\alpha}{\bf{S}}_i\cdot {\bf{s}}_{i,\alpha}
    -{t^*\over 2\sqrt{d}}\sum_{<ij>,\alpha,\sigma}
    \left(c_{i,\alpha,\sigma}^{\dag} c_{j,\alpha,\sigma}
    +\mbox{h.c.}\right) $$
\begin{equation}
    -\mu\sum_{i,\alpha,\sigma}c_{i,\alpha,\sigma}^{\dag}
    c_{i,\alpha,\sigma}\,,
    \label{Ham}
\end{equation}
where $c_{i,\alpha,\sigma}^{\dag}$ ($c_{i,\alpha,\sigma}^{\dag}$)
creates (destroys) an electron on site $i$ in channel $\alpha=1,2$
of spin $\sigma$, ${\bf{S}}_i$ is the Kondo spin on site $i$, and
${\bf{s}}_{i,\alpha}$ are the conduction electron spin operators
for site $i$ and channel $\alpha$.  The sites $i$ form an
infinite-dimensional
hypercubic lattice.  Hopping is limited to nearest neighbors with
hopping integral $t\equiv t^*/2\sqrt{d}$ and
the scaled hopping integral $t^*$ determines the energy unit and is set
equal to one $(t^*=1)$.  Thus, on each site the Kondo spin mediates spin
interaction between the two different channels.  This problem is non
trivial, and for the region of interest in which $J>0$ and $T\ll J$,
$t^*$ it
is non perturbative.  Clearly some simplifying method which allows for a
solution of the lattice problem in a non-trivial limit is necessary.

\paragraph*{Formalism.}
Such a method was proposed by Metzner and Vollhardt \cite{mevoll} who
observed that the renormalizations due to local two-particle interactions
become purely local as the coordination number of the lattice increases.
More precisely, the irreducible single-particle self energy
$\Sigma_{\vec{k}}(z)$ and therefore the irreducible two-particle self
energy $\Gamma_{\vec{k},\vec{k}+\vec{q}}(z,z')$ both become independent of
momentum for large coordination number ($2d\to\infty$)\cite{muha89,bramiel90}.
A further consequence of this is that the solution of most standard lattice
models may be mapped onto the solution of a local correlated system coupled
to an effective bath that is self-consistently 
determined\cite{bramiel90,janis,kimkura,volljan,jarrell91,georges92}.  We 
refer the reader to the above references and recent reviews for further
details on the method\cite{infdrev}.  
  
\paragraph*{Simulation.}
In order to solve the remaining impurity problem, we use the
Kondo impurity algorithm of Fye\cite{fye}, modified to simulate the
two-channel problem\cite{luk}.  In the QMC the problem is cast
into a discrete path formalism in imaginary time, $\tau_l$, where 
$\tau_l=l\Delta\tau$, $\Delta\tau=\beta/L$,
and $L$ is the number of times slices.  The values of $L$ used
ranged from $8$ to $96$, with the largest values of $L$
reserved for the largest values of $\beta$  since the time required by the
algorithm increases like $L^3$.  A sign problem was also encountered in
the QMC process and it also limited how low in temperature we could
go.

The Euclidean-time QMC results for the local greens function $G(\tau)$
were then analytically continued to real frequencies using the "annealing"
Maximum Entropy method\cite{jarrell_mem}.  Here, we start from an exact
result for the density of states (DOS) at high temperatures (i.e. a gaussian)
and use this as a default model to analytically continue the highest
temperature
data.  The output is then used as a default model to analytically
continue
the data for the next lower temperature, and so on.  As the temperature
is lowered, only the low-energy features change, so very little
additional
information needs to be added at each temperature.  To ensure that
we have the best possible results from this procedure, we systematically
improve the precision of the QMC data (by running longer) until
the continued spectra stabilize.  The single particle self energy
may then be obtained by inverting the relation
$G(\omega)=-i\sqrt{\pi} w\left(\omega+\mu-\Sigma(\omega)\right)$,
where $w(z)$ is the complex Fadeev function.
 
Most of our simulations were limited to the model at half filling of the
conduction band ($N=1.0$ for $J=0.75, 0.625, 0.5, 0.4$); however, we
have also studied the system away from half filling ($N=0.75$ and $0.50$)
for two values of $J$ ($J=0.75$ and $0.625$).  The qualitative features of
the single-particle properties of the model do not depend strongly upon
filling.
  
\paragraph*{Results.}  As the temperature is lowered, the local
susceptibilities show evidence of Kondo screening: the screened 
local Kondo moment $T\chi(T)$ falls like $T\ln(T)$, and the local 
Kondo spin-conduction spin correlation indicates an antiparallel 
alignment $\left< {\bf{S}}_i\cdot {\bf{s}}_{i,\alpha} \right> <0$; whereas
local spin-spin correlation function between the channels indicates a weak
parallel alignment $\left< {\bf{s}}_{i,2} \cdot {\bf{s}}_{i,1}
\right> >0$.  By comparing the susceptibility of the local moment to 
the impurity susceptibility, we are able to estimate the Kondo scale $T_0$
from the defining relation $T_0\chi(T_0)=0.3$.  We find that 
$T_0 \simeq 0.85 J \exp(-1.01 t^*/J)$.  For comparison, in the impurity 
limit we would anticipate $T_0^{imp} \simeq
\pi^{-1/2} J\exp(-\sqrt{\pi}t^*/J)$.  Hence, $T_0$ is collectively 
enhanced in the lattice, as found in the 
one channel Anderson lattice problem near half 
filling\cite{gutz,grewe,jarrellandy}; we discuss 
our enhancement further below.  
    
When the temperature is lowered below $T_0$, we find non-fermi
liquid behavior in the single-particle properties of the model.  For
example, in Fig.~1 the derivative of the particle distribution function
$n(\epsilon_k)=T\sum_n G(\epsilon_k,i\omega_n)$ is plotted.  This
saturates to a finite width distribution, in contrast to a Fermi liquid
where it would display a dominant delta function contribution as $T\to 0$, 
and a Luttinger liquid\cite{lutnfl} or Marginal fermi liquid\cite{ffnj} 
where it would have a singular divergence.  Clearly we see no such features 
in Fig.~1.  

In Fig. 2 we show one electron properties of the model. 
Fig.~2(a) displays the single particle DOS.  This has a finite value as
$\omega,T\to 0$, with a peak away from the Fermi energy.  
Note that suppression of $N(\omega=0)$ leads to 
an enhanced effective medium DOS which accounts qualtitatively for
our collectively enhanced $T_0$ values.

Novel behavior is 
seen in the real part of the one electron self energy which has positive 
slope at $\omega\to 0$ (Fig.~2(c)).  For a Fermi
liquid, this slope would be negative.  The physical content is
important: $Z=1/(1-\partial Re\Sigma/\partial \omega)$ measures the
overlap of the quasiparticle wave function with the original
one-electron wave function having the same quantum labels. A positive
slope leads to $Z>1$ or $Z<0$, indicating a breakdown of the quasiparticle
concept.  Concommitant with the finite width of $-dn/d\epsilon_k$ is a 
finite imaginary part to the low temperature self energy (Fig.~2(b)).  
This indicates that the one-electron excitations are ill defined on approach 
to the Fermi surface, again ruling out a Fermi liquid description.   Since 
the low temperature thermodynamic properties such as the specific heat, 
uniform magnetic susceptibility, and charge susceptibility display no evidence 
for a gap, we believe the observed behavior indicate a new kind of metallic 
state.  

From our numerical results, one can
show that the zero temperature self energy must be non-analytic.  If the
self energy is analytic everywhere on the real axis (and the upper half
complex plane), then one may easily show that
\begin{equation}
\lim_{T\to 0} \frac{{\rm{Im}}\Sigma(i\omega_0)}{\omega_0}=
\lim_{T\to 0}
\left.\frac{d{\rm{Re}}\Sigma(\omega)}{d\omega}\right|_{\omega=0}\, ,
\label{analsig}
\end{equation}
where $\omega_0=\pi T$ is the lowest fermionic Matsubara frequency
along the imaginary axis. However, we find that when $T<T_0$,
$\frac{{\rm{Im}}\Sigma(\omega_0)}{\omega_0}<0$ and is apparently divergent;
whereas, from Fig.~3 it is apparent that
$\left.\frac{d{\rm{Re}}\Sigma(\omega)}{d\omega}\right|_{\omega=0} > 1$.
Thus, $\Sigma(\omega)$ cannot be analytic at the origin of the complex
plane.
Given this, and the apparent shape of $\Sigma(\omega)$ in
Fig.~2(b), we postulate the form ${\rm{Im}}\Sigma(\omega)\sim -c +
|\omega|$.
 
The non-fermi liquid behavior also strongly effects 
experimentally relevant transport properties of the system.  
The electrical resistivity is shown in Fig.~3.  
We find that $\rho(T)/\rho(0)$ curves for different $J$ values
collapse onto a universal scaling curve when 
plotted against $T/T_0$.  As shown in the inset, $\rho(T)\simeq
\rho(0)[1+B(T/T_0)]$ for $T\to 0$, with $B<0$\cite{caveat}. 
We interpret the finite value of $\rho(0)$  
together with $-Im\Sigma(0,0)$ as ``spin
disorder scattering'' off of the degenerate screening clouds
centered about each local moment spin.  Given the finite density of one
particle excitations at the Fermi energy, this finite residual
resistivity is indicative of an ``incoherent metal phase'' brought about
by the disordered spin degrees of freedom, in qualitative agreement with
results obtained with a Lorentzian bare conduction DOS (which doesn't
self-consistently renormalize)\cite{coxinfd}.  We conjecture that 
the antiferromagnetism at half filling will lift the disorder at each
site and produce an insulating state due to the cell doubling.

We have also computed the optical conductivity (not shown) and find that
it displays vanishing Drude weight at low temperature together with a
finite frequency peak.  Both these features again support our
interpretation in terms of a new kind of non-Fermi liquid metallic
state. 
 
We find a commensurate antiferromagnetic transition at half filling for the
simple nearest neighbor hopping model.  We can suppress this phase by
adding next neighbor hopping along each principle axis which will not alter 
our conclusions for the paramagnetic phase.  The antiferromagnetic order is
suppressed away from half filling.  The complete phase diagram of the
model will be explored in a separate publication.  

Finally, we mention the possible applicability of our results to
concentrated heavy fermion systems.  Three systems display resistivity 
of the form $\rho(T)\approx \rho(0)[1+B(T/T_0)^{\alpha}]$ for $T<T_0$, 
with $\rho(0)$ of order the unitarity limit and $\alpha \approx 1$.
They are: 
UCu$_{5-x}$Pd$_x$ (with $B<0$)\cite{hfnfl}, UBe$_{13}$ ($B>0$)\cite{stegnew},
and CeCu$_2$Si$_2$ ($B>0$)\cite{steggoa}.  While UCu$_{5-x}$Pd$_x$ is
a possible example of the distribution of Kondo scales
scenario\cite{disknfl}, UBe$_{13}$ and CeCu$_2$Si$_2$ 
are ordered compounds which have been proposed as
possible two-channel lattice systems (see Refs.\cite{twochnfl}(b,c));  
it is conceivable that $1/d$ corrections can effect the sign change of 
$B$ relative to our results.  Another key difference with our model is
that we assume {\it global} $SU(2)|_{spin}\times SU(2)|_{channel}$
symmetry, while in the real materials these can only be local
symmetries.

We would like to acknowledge useful discussions with F.\ Anders, W.\ Chung,
A.\ Georges, M.\ Ma, A.J.\ Millis, 
and W.\ Putikka.  Jarrell and Pang would like to acknowledge 
the support of NSF grants DMR-9406678 and DMR-9357199. Cox acknowledges
the support of the U.S. Department of Energy, Division of Basic Energy
Sciences, Office of Materials Research, and, at the ITP, by NSF Grant
No. PHY94-07194. Computer support was provided by the Ohio 
Supercomputer Center.

\begin{figure}
\epsfxsize=3.8in
\epsffile{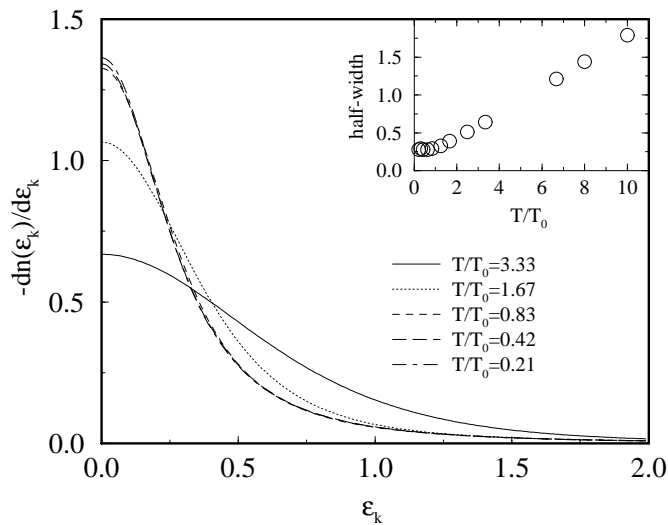}
\caption{Derivative of the particle distribution function
$n(\epsilon_k)=T\sum_n G(\epsilon_k,i\omega_n)$ when $J=0.625$ and $N=1.0$.
In a non-interacting system or one with a fermi-liquid low temperature
state, $-dn(0)/d\epsilon_k\propto Z^2\beta$ and the distribution should have
a width $\propto T$ which is the width of the fermi edge and is
proportional
to the scattering rate at the fermi surface.  However, in this case
$-dn(0)/d\epsilon_k$ saturates at low T and so does the width of
$-dn(\epsilon_k)/d\epsilon_k$ , shown in the inset, indicating that a
fermi liquid has not formed and that scattering rate remains finite
when $\epsilon_k=\mu$ as $T\to 0$.}
\end{figure}

\begin{figure}[t]
\epsfxsize=3.8in
\epsffile{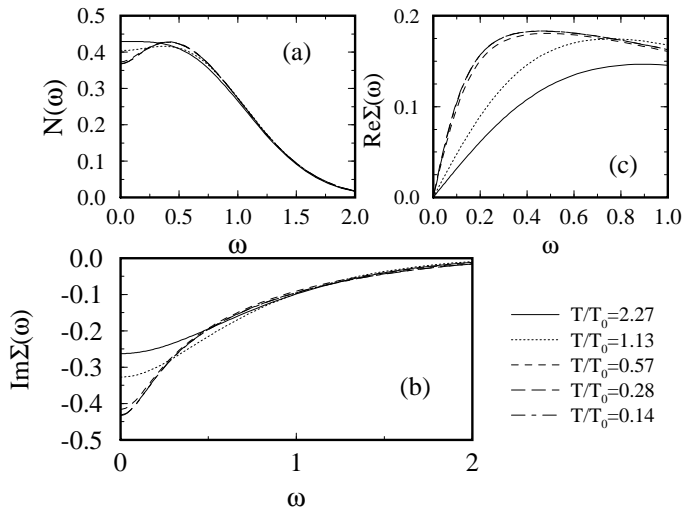}
\caption{Single particle properties of the two-channel Kondo lattice in
infinite dimensions when $J=0.750$ and $N=1.0$.  (a) Single-particle density 
of states (DOS).  At high temperatures $T\gg T_0$ (not shown), the DOS is 
a gaussian, crossing over to the peaked distribution with relative suppression  
at $\omega\to 0$ for lower temperatures when $T\ll T_0$.  (b) Imaginary part 
of the self energy.  As the temperature is lowered the self energy does not 
approach a fermi liquid form ${\rm{Im}}\Sigma(\omega)\propto -T^2-\omega^2$, 
but rather appears to approach the non-analytic form (see text)
${\rm{Im}}\Sigma(\omega)\propto -c+\left|\omega\right|$.  (c) The real
part of the self energy ${\rm{Re}}\Sigma(\omega)$, is also anomalous since 
its initial slope is positive indicating a quasiparticle renormalization 
factor which is greater than one.}

\end{figure}

\begin{figure}[t]
\epsfxsize=3.8in
\epsffile{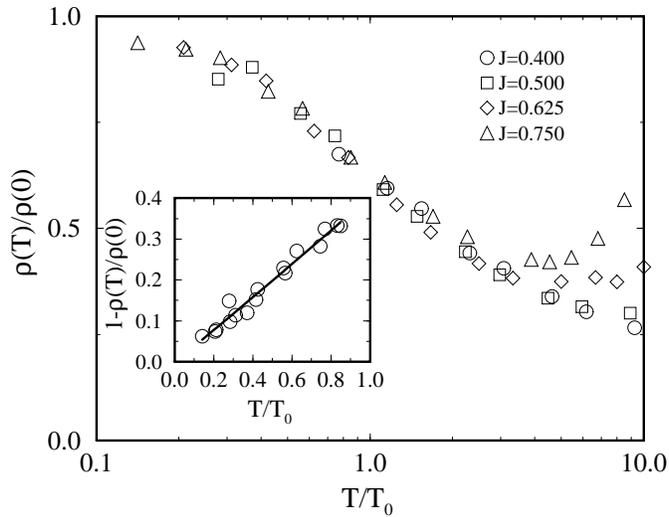}
\caption{Resistivity of the two-channel Kondo lattice.  $\rho(T)/\rho(0)$ is 
plotted versus $T/T_0$ for various values of $J$.  In the inset, the lowest 
temperature data (for $T/T_0 < 1$, shown as open circles) was fit 
to $\rho(T)/\rho(0)=1+B\left( T/T_0\right)^\alpha$, with $B=-0.4$ and 
$\alpha=1.03$. }
\end{figure}


\begin{references}
\bibitem{caveat} We note that we cannot rule out the appearance of a 
different power law at temperatures lower than $T/T_0=0.14$.
\end{references}
\end{document}